\documentclass{article}

\usepackage[usenames,dvipsnames,table]{xcolor}   
\usepackage{graphicx} 
\usepackage{arxiv}
\usepackage{times}
\usepackage{hyperref}
\usepackage{breakcites}
\usepackage{inconsolata}
\usepackage{enumitem}
\usepackage{notation}
\usepackage{cleveref}
\usepackage{xspace}
\usepackage{pdfpages}
\usepackage[T1]{fontenc}

\newcommand{\LiteEFG}{{\tt LiteEFG}\xspace}
\newcommand{\CF}{\text{CF}}

\title{\LiteEFG: An Efficient Python Library for Solving Extensive-form Games}
\author{Mingyang Liu $^1$, Gabriele Farina$^1$, Asuman Ozdaglar $^1$\\
    $^1$ LIDS, EECS, Massachusetts Institute of Technology\\
    $^1$ \texttt{\{liumy19,asuman,gfarina\}@mit.edu}\\
}

\begin{document}
\maketitle

\begin{abstract}
    \LiteEFG is an efficient library with easy-to-use Python bindings, which can solve multiplayer extensive-form games (EFGs). \LiteEFG enables the user to express computation graphs in Python to define updates on the game tree structure. The graph is then executed by the C++ backend, leading to significant speedups compared to running the algorithm in Python. Moreover, in \LiteEFG, the user needs to only specify the computation graph of the update rule in a decision node of the game, and \LiteEFG will automatically distribute the update rule to each decision node and handle the structure of the imperfect-information game.
\end{abstract}

\begin{center}
\begin{minipage}{.89\textwidth}
\tableofcontents
\end{minipage}
\end{center}

\newpage
\section{Introduction}

The successes of reinforcement learning in solving various games, including Go \citep{silver2016mastering, silver2017mastering}, Atari \citep{mnih2013playing-atari}, and Dota2 \citep{berner2019dota}, have increased interest in developing scalable approaches for finding equilibrium strategies in \emph{extensive-form games (EFGs)}. Compared to Markov games where the game state is fully observable, a big challenge in EFG is that decisions are made according to partial observation of the game state. 
It further results in additional hardness in computing the expected utility of taking an action, unlike Q-values in Markov games, since the expected utility depends on the distribution of hidden information. 
For instance, in Texas Hold'em, the utility of raising the bid depends not only on the private hands of the decision maker but also on those of her opponents, which are unknown when making the decision.

The recent interest in computation methods for solving EFGs also sparked activity in developing libraries for executing algorithms over the game tree. A popular library is OpenSpiel \citep{LanctotEtAl2019OpenSpiel}, which provides various game environments with an easy-to-use Python API for researchers to test their algorithms. However, the algorithms operating over OpenSpiel are usually very slow since they are executed via Python. This motivates the need to devise a library that incorporates both simple Python API and efficient backend execution, similar to how TensorFlow \citep{abadi2016tensorflow} and PyTorch \citep{steiner2019pytorch} operate.


To address the challenge above, we propose \LiteEFG, an open-source library with a simple Python API and an efficient C++ backend for solving EFGs, which is much faster than pure Python execution. With \LiteEFG, researchers need to define the update-rule at a decision node in a similar way as defining neural networks with TensorFlow or Pytorch, then \LiteEFG will automatically distribute that update-rule to each individual decision node and handle the relationship between different decision nodes automatically. Compared to OpenSpiel \citep{LanctotEtAl2019OpenSpiel}, {\tt \LiteEFG} is simpler and faster when solving tabular games (games that may fit into the computer memory). In experiments, the classical baseline, Counterfactual Regret Minimization \citep{DBLP:conf/nips/ZinkevichJBP07-CFR}, implemented by \LiteEFG is about $100\times$ faster than that of OpenSpiel.

In \LiteEFG, researchers only need to specify the computation graph of the algorithm via Python. Then, the computation graph will be executed via C++, which provides acceleration by several orders of magnitude. Moreover, due to the imperfect information of EFG, the game states and decision nodes no longer coincide, which complicates the implementation of the algorithm. To simplify the issue, \LiteEFG will automatically aggregate the information from different game states belonging to the same decision node \footnote{In EFGs, since the game is partially observable, some game states may not be differentiable for the decision maker. For instance, in Texas Hold'em, the game states that only differ in the private hands of the opponents are not differentiable for the decision maker. Then, we call those game states belonging to the same decision node, because the decision made at those game states should be the same.}, so users only need to specify the update-rule for the decision node, without concerning with the aggregation process.

\section{Preliminaries}
In this section, we will introduce the preliminaries of EFGs.
%
%
We use $\Delta^m\coloneqq\cbr{\bx\in[0,1]^m\colon \sum_{i=1}^m x_i=1}$ to denote the $m-1$ dimensional probability simplex. For a discrete set $\cC$, we use $|\cC|$ to denote its cardinality. For any real number $x\in\RR$, $[x]^+\coloneqq x\cdot \ind_{x\geq 0}$, which is $x$ when $x\geq 0$ and $0$ otherwise.

\paragraph{Basics of extensive-form games.} In an $N$-player EFG, we use $[N]\coloneqq \cbr{1,2,...,N}$ to denote the set of all players. Optionally, a fictitious player---called the \emph{chance player}---is introduced to model stochastic events, such as a random draw of cards or dice roll.

The game is a tree structure and we use $\cH$ to denote the set of all nodes in the game. For each $h\in\cH$, one of the players among $\cbr{c}\cup [N]$, where $c$ is the chance player, will take actions at $h$. We use $p(h)$ to denote the player acting at $h$, and say that ``$h$ belongs to $p(h)$''. Then, player $p(h)$ can choose an action in the action set $\cA_h$ and the process will be repeated until the game reaches a terminal node $h'\in\cZ$, where $\cZ$ is the set of all terminal nodes. The utility for each player $i\in [N]$ is denoted by $\cU_i : \cZ\to [0, 1]$. 

\paragraph{Information Set.} In an imperfect-information game, a player $i\in[N]$ may not be able to distinguish all the nodes of the tree, that is, $\cbr{h^1,h^2,...,h^k\colon \forall j=1,2,...,k, p(h^j)=i}$. For example, in the two-player Texas Hold'em, each player cannot distinguish nodes that only differ because of her opponent's private hand, since the hand of the opponent is not revealed. To model imperfect information, a partition of each player's nodes---called the player's \emph{information partition}---is introduced. The elements of the partition, called \emph{information set} (or infoset for short), denote nodes that are indistinguishable to the player when acting at any of them. $\cS_i$ is used to denote the set of all infosets of player $i\in[N]$. For simplicity, for each node $h$, we use $s(h)$ to denote the infoset that $h$ belongs to and extend the player indicator $p$ to infosets, \emph{i.e.} $p(s(h))\coloneqq p(h)$.

\paragraph{Strategy of Players.} The strategy of player $i\in[N]$ can be written as $\pi_i(\cdot\given s)\in\Delta^{|\cA_s|}$ for each infoset $s\in\cS_i$, where $\cA_s$ is the action set of infoset $s$.\footnote{Since nodes in $s$ cannot be differentiated by $i$, the action set of each node in $s$ must be the same.} Therefore, infosets are also called \emph{decision nodes}, since players make decisions conditioned on the infoset.

\paragraph{Reach Probability and Sequence-form Strategy.} For any strategy $\pi_i$ of player $i\in[N]$, we can define the reach probability of player $i$ as $\mu_i^{\pi_i}(h^1\to h^2)$ as the reach probability from node $h^1$ to node $h^2$, to which we only count the probability contributed by player $i$ while applying $\pi_i$. Moreover, we can define $\mu_i^{\pi_i}(s\to h)$ as the reach probability from infoset $s\in\cS_i$ to node $h$, when only counting the probability contributed by $\pi_i$. We use $\mu_i^{\pi_i}(\emptyset\to h)$ ($\mu_i^{\pi_i}(\emptyset\to s)$) as the reach probability to $h$ ($s$) from the root of the game. With reach probability, we slightly abuse the notion $\mu$ to denote the sequence-form strategy $\mu_i^{\pi_i}(s, a)\coloneqq \mu_i^{\pi_i}(\emptyset \to s)\pi_i(a\given s)$. 

\section{Tour: Implementation of Counterfactual Regret Minimization (CFR)}

In this section, we will introduce how to use \LiteEFG.

\subsection{Basics of Computation Graph}

\LiteEFG is based on the computation graph, in which a vector is stored at each node and users need to define the relationships between graph nodes. For instance, node $A$ equals node $B$ plus node $C$. Then, every time the user updates the graph, the variables at each graph node will be updated according to the predefined relationship. 

In \LiteEFG, the user need to define the computation graph for an infoset first. Then, the graph will be copied to each infoset in $\bigcup_{i\in[N]}\cS_i$. Therefore, all infosets share the same relationship between graph nodes, while the variables stored in the graph node of each infoset are independent.

In \LiteEFG, the user can define a node by {\tt LiteEFG.function(...)} with some function of \LiteEFG\footnote{The full API list can be found in \url{https://github.com/liumy2010/LiteEFG}.}, such as {\tt LiteEFG.sum and LiteEFG.exp}. In this case, \LiteEFG will create a new node to store the outcome of the function and return that node.
Alternatively, the user can update the variable at a node by {\tt x.inplace(LiteEFG.function(...))}. In this case, \LiteEFG will {\bf not} create a new node. Instead, the outcome of the function will be stored at node {\tt x}, and replace the original variable at {\tt x}.

\subsection{Construction of Computation Graph}

In this section, we will introduce the construction of the computation graph for Counterfactual Regret Minimization (CFR) \citep{DBLP:conf/nips/ZinkevichJBP07-CFR}, one of the most prominent algorithms for solving EFGs, with \LiteEFG.

\subsection{Visitation of Infosets}

\begin{figure}[t]
    \centering
    \includegraphics[width=0.9\linewidth]{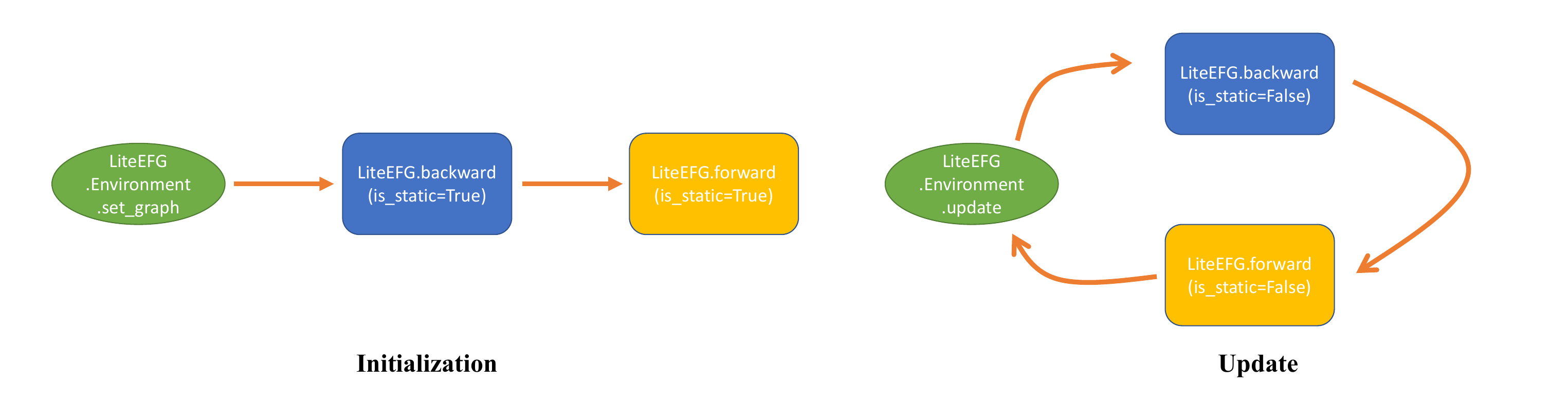}
    \caption{At the very beginning, when the computation graph for the environment is determined by {\tt LiteEFG.Environment.set\_graph}, the static backward nodes and static forward nodes will be executed sequentially. Then, each time {\tt LiteEFG.Environment.update} is called, the dynamic backward nodes and dynamic forward nodes will be executed.}
    \label{fig:order-of-graph}
\end{figure}

\LiteEFG provides four types of graph nodes.
\begin{itemize}
    \item {\tt LiteEFG.backward(is\_static=True)}: Static backward nodes. These nodes will be executed at initialization (ahead of the execution of any other nodes). To execute the static backward nodes, infosets will be visited in the reversed breadth-first order and the corresponding static backward nodes will be executed.
    \item {\tt LiteEFG.forward(is\_static=True)}: Static forward nodes. These nodes will be executed at initialization (ahead of the execution of any dynamic nodes, but after the static backward nodes). To execute the static backward nodes, infosets will be visited in the breadth-first order and the corresponding static forward nodes will be executed.
    \item {\tt LiteEFG.backward(is\_static=False)}: Dynamic backward nodes. These nodes will be executed every time the function {\tt LiteEFG.Environment.update} is called. To execute the dynamic backward nodes, infosets will be visited in the reversed breadth-first order and the corresponding dynamic backward nodes will be executed.
    \item {\tt LiteEFG.forward(is\_static=False)}: Dynamic forward nodes. These nodes will be executed every time the function {\tt LiteEFG.Environment.update} is called. To execute the dynamic backward nodes, infosets will be visited in the breadth-first order and the corresponding dynamic forward nodes will be executed.
\end{itemize}

The order is also illustrated in \Cref{fig:order-of-graph}.

\subsubsection{Static Graph}

\begin{figure}[t]
    \centering
    \includegraphics[width=0.9\linewidth]{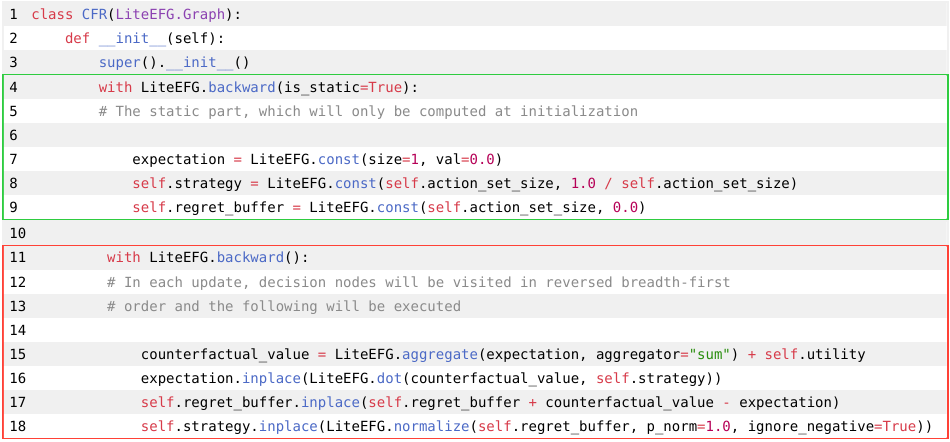}
    \caption{Implementation of Counterfactual Regret Minimization (CFR) \citep{DBLP:conf/nips/ZinkevichJBP07-CFR} by \LiteEFG. The green box displays the definition of static variables that will only be updated once at initialization. The red box displays the variables that will be updated every time updating the graph.}
    \label{fig:API-CFR}
\end{figure}

The update-rule of CFR is as follows. For any player $i\in[N]$, an infoset $s\in\cS_i$, and action $a\in\cA_s$, the update-rule at timestep $t\in \cbr{1,2,...,T}$ is
\begin{align}
    &\CF^{(t+1)}_i(s,a)= \sum_{h\in\cZ} \cU_i(h) \frac{\mu_i^{\pi_i^{(t+1)}}(s\to h)}{\pi_i^{(t+1)}(a\given s)}\prod_{j\in\cbr{c}\cup[N]\colon j\not=i} \mu_j^{\pi_{j}^{(t+1)}}(\emptyset\to h) \notag\\
    &R_i^{(t+1)}(s,a)=R_i^{(t)}(s,a)+ \CF_i^{(t+1)}(s,a) -\EE_{a'\sim \pi^{(t+1)}_i(\cdot\given s)}\sbr{\CF_i^{(t+1)}(s,a')}\label{eq:update-rule-CFR}\\
    &\pi^{(t+2)}_i(a\given s)=\begin{cases}
        \frac{[R_i^{(t+1)}(s,a)]^+}{\sum_{a'\in\cA_s} [R_i^{(t+1)}(s,a')]^+}&\sum_{a'\in\cA_s} [R_i^{(t+1)}(s,a')]^+>0\\
        \frac{1}{|\cA_s|}&\sum_{a'\in\cA_s} [R_i^{(t+1)}(s,a')]^+=0
    \end{cases}\notag
\end{align}

From \Cref{eq:update-rule-CFR}, to update CFR, we need to maintain two variables in each infoset $s$, the strategy $\bpi_i(\cdot\given s)\in\Delta^{|A_s|}$, and the regret buffer $R(s,\cdot)\in\RR^{|A_s|}$. In \Cref{fig:API-CFR}, the \textcolor{green!60!black}{green code block} defines the static backward nodes of CFR algorithm. In line 7, we define {\tt expectation}, which is the placeholder for $\EE_{a'\sim \pi^{(t+1)}_i(\cdot\given s)}\sbr{\CF_i^{(t+1)}(s,a')}$. We will discuss the necessity of such placeholder in \Cref{sec:aggregate}. Line 8 and 9 initialize $\bpi_i(\cdot\given s)$ as uniform distribution over $\Delta^{|\cA_s|}$ and $R(s,\cdot)$ as zero vector individually. The vector at {\tt self.action\_set\_size} is a scalar equivalent to $|\cA_s|$ for each infoset $s$.

\subsubsection{Dynamic Graph}
\label{sec:aggregate}

In this section, we will focus on the {\color{red!80!black} red code block} in \Cref{fig:API-CFR}.

\paragraph{Line 15.}
Line 15 displays the usage of {\tt LiteEFG.aggregate}.

\begin{align*}
    {\tt LiteEFG.aggregate(}&{\tt x, aggregator\_name : ["sum", "mean", "max", "min"],}\notag\\
    &{\tt object\_name : ["children", "parent"]="children",}\\
    &{\tt player : ["self", "opponents"]="self", padding=0.0)}
\end{align*}

\begin{itemize}
    \item {\tt x}: Indicates the graph node from which the function aggregates information.
    \item {\tt aggregator\_name}: Method to aggregate the information.
    \item {\tt object\_name}: Aggregate information from parent / children of the current infoset. By default, it is ``children".
    \item {\tt player}: Aggregate information from the infosets belong to self / opponents. For instance, when {\tt object\_name="children"}, for an infoset $s\in\cS_i$, we will aggregate the information from the children $s'$ of $s$, with $p(s')=i$ if \verb|player="self"| and $s'$ with $p(s')\not=i$ otherwise.
    \item {\tt padding}: If the parent / children of the current infoset does not exist, return {\tt padding}.
\end{itemize}

When the {\tt object\_name} is ``children'', for each action $a\in\cA_s$ in infoset $s$, there may be several subsequent children infoset. {\tt aggregate} will first concatenate the variable at {\tt x} at all those infosets together. Then, \LiteEFG will call the aggregator function specified in {\tt aggregator\_name} to aggregate that vector to a single scalar. Finally, \LiteEFG will concatenate the scalar at each action $a\in\cA_s$ to a vector in $\RR^{|\cA_s|}$.

When the {\tt object\_name} is ``parent'', suppose $(s',a')$ is the parent sequence of $s$. If the variable $\bv$ stored at {\tt x} at $s'$ is in $\RR^{|\cA_{s'}|}$, then $v_{a'}$ will be returned. Or if $\bv$ will be returned if it is a scalar. Otherwise, an error occurs.

\paragraph{Line 16.} Line 16 computes $\EE_{a'\sim \pi^{(t+1)}_i(\cdot\given s)}\sbr{\CF_i^{(t+1)}(s,a')}$ to replace the old value $\EE_{a'\sim \pi^{(t)}_i(\cdot\given s)}\sbr{\CF_i^{(t)}(s,a')}$ stored in {\tt expectation}. Because we need to specify {\tt x} as {\tt expectation} for the aggregator function, while {\tt expectation} is defined upon the returned value of the {\tt aggregate} function, we have to put a placeholder of {\tt expectation} at line 7 before {\tt aggregate}.

\paragraph{Line 17.} Line 17 updates the value of $R^{(t)}_i(s,\cdot)$ to $R^{(t+1)}_i(s,\cdot)$ according to \Cref{eq:update-rule-CFR}.

\paragraph{Line 18.} Line 18 computes $\bpi_i^{(t+2)}(\cdot\given s)$ according to \Cref{eq:update-rule-CFR} to replace the old value $\bpi_i^{(t+1)}(\cdot\given s)$ stored in {\tt self.strategy}.

\subsection{Loading the Game}

\begin{figure}[t]
    \centering
    \fbox{\includegraphics[width=0.9\linewidth]{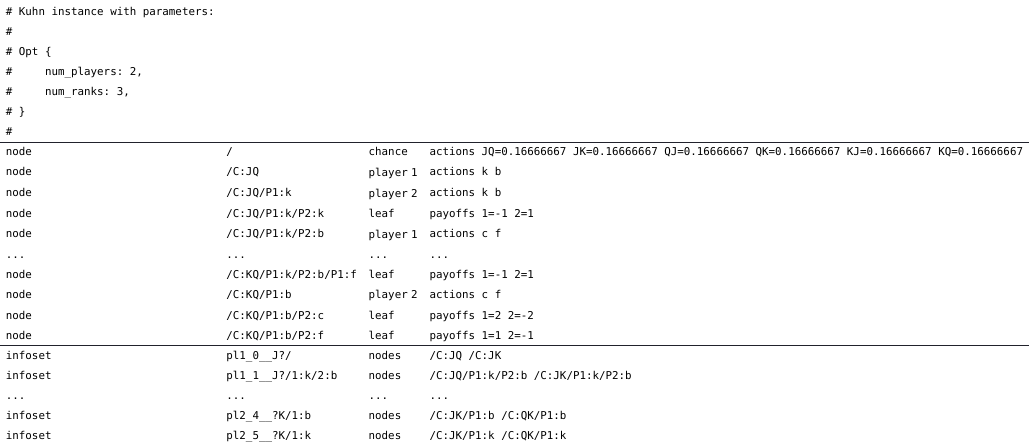}}
    \caption{An example of Kuhn Poker \citep{Kuhn} represented in the game file format supported by \LiteEFG.}
    \label{fig:game-file-example}
\end{figure}

\LiteEFG is fully compatible with OpenSpiel, \emph{i.e.} \LiteEFG supports almost all games in OpenSpiel\footnote{Some games such as Hanabi \citep{bard2020hanabi} is too big and the implementation of OpenSpiel does not include infoset, so that \LiteEFG does not support it.}. Moreover, for games not implemented by OpenSpiel, users can write a game description text file alternatively and load it using {\tt LiteEFG.FileEnv}. An example of the game file is illustrated in \Cref{fig:game-file-example}, and the full example can be found in {\tt LiteEFG/game\_instances/kuhn.game}. At the beginning, the game file displays the parameters of the game, where the parameter {\tt num\_players} is necessary and other parameters are optional. 
The next several lines will include the node information, with an identifier {\tt node} at the beginning of the line, and the node's name goes after it. For different types of nodes, the additional information should obey the following rules,
\begin{itemize}
    \item {\bf Chance Node} ($p(h)=c$): Keyword {\tt chance} should be placed at first and the chance events leading by {\tt actions} will be displayed afterward. The format is {\tt event\_name=probability}. For instance, in \Cref{fig:game-file-example}, the chance event {\tt JQ} means the private card dealt to player 1 and 2 is Jack and Queen individually.
    \item {\bf Player Node} ($p(h)\in [N]$): Keyword {\tt player} should be placed first and $p(h)$ should go after it. Then, the valid actions leading by {\tt actions} should be placed afterward. In \Cref{fig:game-file-example}, {\tt k, c, f, b} is check, call, fold, and bet individually.
    \item {\bf Leaf (Terminal) Node} ($h\in\cZ$): Keyword {\tt leaf} should be placed first and the utility leading by leading by {\tt payoffs} go afterward. The format of utility is $i=\cU_i(h)$, where $i\in[N]$ is the player index.
\end{itemize}
After that, the infosets will be described. Each line of infoset description will be led by the identifier {\tt infoset} and the name of the infoset will be placed after it. Then, the nodes in that infoset will be introduced. After the keyword {\tt nodes}, the name of those nodes in the infoset will be displayed.

\subsection{Training}

\begin{figure}[t]
    \centering
    \includegraphics[width=0.9\linewidth]{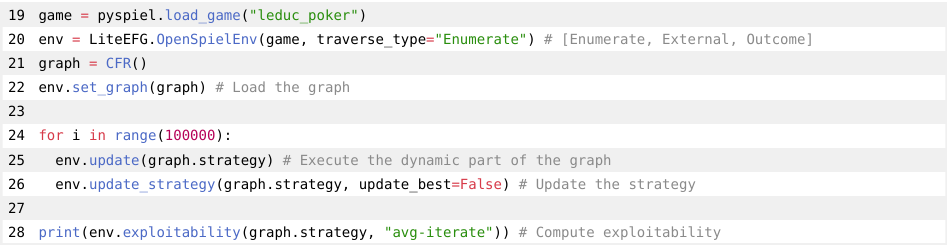}
    \caption{Training and evaluation of CFR algorithm defined in \Cref{fig:API-CFR}.}
    \label{fig:training-eval}
\end{figure}

In line 19 and line 20 of \Cref{fig:training-eval}, we load the environment {\tt leduc\_poker} from OpenSpiel to \LiteEFG. The {\tt traverse\_type} specifies how to traverse the game tree in each iteration. Currently, \LiteEFG supports enumerating all nodes, external sampling\footnote{For each player $i\in [N]$, we traverse the game tree once according to the following rule. For a node with $p(h)=i$, we visit children under $(h,a)$ for each $a\in\cA_h$. For a node with $p(h)\not=i$, we sample a children $a\sim \pi_{p(h)}(\cdot\given s(h))$ and only visit the children under $(h,a)$.} \citep{lanctot2009monte-carlo-cfr}, and outcome sampling\footnote{Whenever meets a node, sample a children $a\sim \pi_{p(h)}(\cdot\given s(h))$ and only visit the children under $(h,a)$.} \citep{lanctot2009monte-carlo-cfr}. 

In line 21 and line 22 of \Cref{fig:training-eval}, we load the computation graph of CFR defined in \Cref{fig:API-CFR} into the environment. In line 25, we update the dynamic part of the graph, so that the update-rule in \Cref{eq:update-rule-CFR} is executed.

CFR algorithm only guarantees the average sequence-form strategy will converges to the Nash equilibrium in a two-player zero-sum game, that is
\begin{align}
    \rbr{\frac{1}{T}\sum_{t=1}^T \mu_1^{\pi_1^{(t)}}, \frac{1}{T}\sum_{t=1}^T \mu_2^{\pi_2^{(t)}} } \text{ converging to the Nash equilibrium.}
\end{align}
Therefore, we need to maintain the average-iterate sequence-form strategy. To simplify the implementation, \LiteEFG provides the function {\tt LiteEFG.Environment.update\_strategy}, which will automatically maintain the average-iterate sequence-form strategy (also supports other types of sequence-form strategy, and the details can be found in the Github repository).

\subsection{Evaluation}

In line 28 of \Cref{fig:training-eval}, we print the exploitability of the strategy stored at graph node {\tt graph.strategy}. {\tt "avg-iterate"} indicates we want to measure that of the average-iterate strategy. {\tt LiteEFG.Environment.exploitability} returns a vector $\bv$, where each element $v_i\geq 0$ indicates how much player $i$ can improve her utility by deviating to other strategies while the strategies of other players remain fixed. The sum $\sum_{i\in[N]} v_i$ is the exploitability, which measures the distance to the Nash equilibrium. 
When the exploitability is zero, it implies that the current strategies of all players form a Nash equilibrium.

\subsection{Debug}

\begin{figure}[h]
    \centering
    \includegraphics[width=0.99\linewidth]{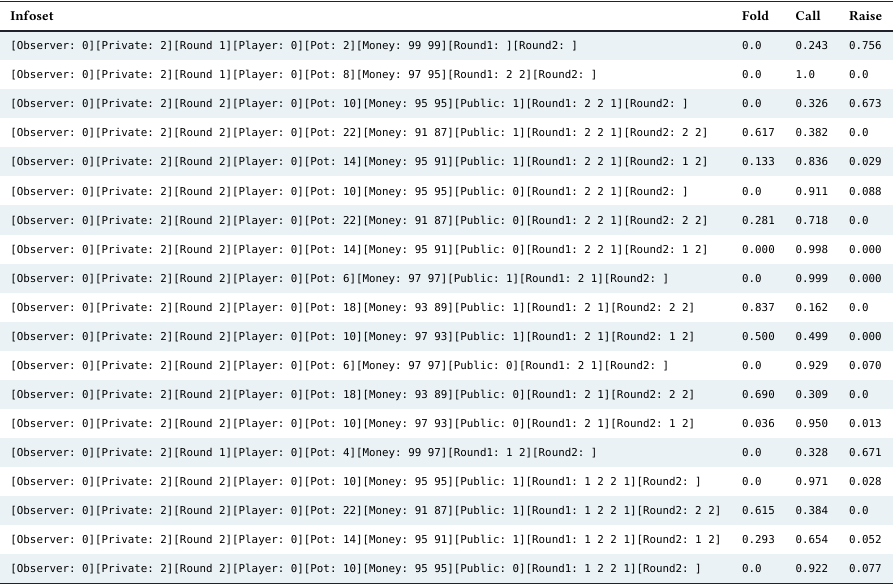}
    \caption{A snapshot of the strategy of player 1 generated by CFR algorithm in {\tt leduc\_poker(suit\_isomorphism=True)} of OpenSpiel. The first column displays the name of the infoset, and the second to the fourth column is the probability of choosing fold / call / raise in that infoset. The index of fold / call / raise in the representation of infoset is 0, 1, 2.}
    \label{fig:strategy}
\end{figure}

To debug the strategy, users can call {\tt LiteEFG.OpenSpielEnv.get\_strategy} to get a {\tt pandas.DataFrame} as shown in \Cref{fig:strategy} to display the strategies.

\begin{figure}[h]
    \centering
    \includegraphics[width=0.9\linewidth]{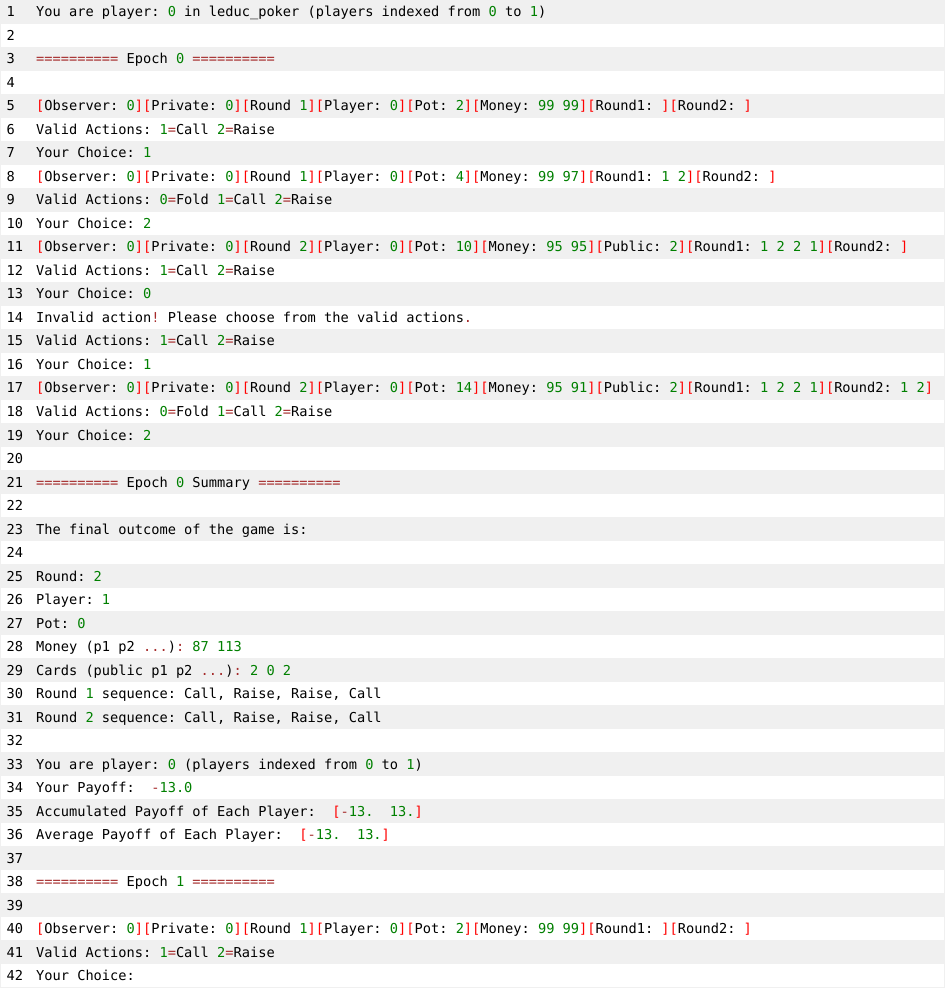}
    \caption{The interaction with strategy generated by CFR in {\tt LiteEFG.OpenSpielEnv.get\_strategy}.}
    \label{fig:interact}
\end{figure}

Moreover, the user can also interact with the strategy computed by the algorithm, by calling {\tt LiteEFG.OpenSpielEnv.interact}. The interaction is displayed in \Cref{fig:interact}.

\subsection{Various Baselines}

Though EFG is developing fast in recent years \citep{DBLP:conf/nips/LeeKL21-EFG-last-iterate,DBLP:conf/iclr/LiuOYZ23-power-reg,DBLP:conf/iclr/SokotaDKLLMBK23-MMD}, it is hard to make a fair comparison between different algorithms. The difficulty is mainly two-fold. Firstly, some algorithms are not open-sourced and algorithms in EFGs are sensitive to the hyper-parameters. Therefore, when re-implementing the baseline algorithms, researchers may not be able to achieve the same performance as the original paper with the same set of parameters. Moreover, sometimes researchers do not use OpenSpiel or even not using Python to implement their algorithms. Secondly, even though some algorithms are open-sourced \citep{DBLP:conf/iclr/SokotaDKLLMBK23-MMD}, they are highly inefficient since they are purely based on Python. As a result, it takes a lot of computation resources and time to make a fair comparison with previous results. Therefore, comprehensive and efficient baselines are important for the whole community, and \LiteEFG provides a variety of baselines. We list the baseline algorithms in \Cref{table:baselines}.

\begin{table}[h]\centering
  \rowcolors{2}{gray!25}{white}
  \renewcommand{\arraystretch}{1.2}
\scalebox{.90}{\begin{tabular}{lp{3cm}l}
\toprule
{\bf Algorithms}                                                    & {\bf Traverse Type}                                                                  & {\bf Reference} \\ \midrule
Counterfactual Regret Minimization (CFR)                      & Full Information                                                               &     \citet{DBLP:conf/nips/ZinkevichJBP07-CFR}      \\ 
Counterfactual Regret Minimization+ (CFR+)                    & Full Information                                                               &      \citet{DBLP:conf/ijcai/TammelinBJB15-CFR+}     \\ 
Discounted Counterfactual Regret Minimization (DCFR)          & Full Information                                                               &    \citet{DBLP:conf/aaai/BrownS19-DCFR}       \\ 
Predictive Counterfactual Regret Minimization (PCFR)          & Full Information                                                               &      \citet{farina2021faster-blackwell}     \\ 
External-Sampling Counterfactual Regret Minimization (ES-CFR) & External Sampling                                                              &      \citet{lanctot2009monte-carlo-cfr}     \\ 
Outcome-Sampling Counterfactual Regret Minimization (OS-CFR)  & Outcome Sampling                                                               &      \citet{lanctot2009monte-carlo-cfr}     \\ 
Dilated Optimistic Mirror Descent (DOMD)                      & Full Information                                                               &      \citet{DBLP:conf/nips/LeeKL21-EFG-last-iterate}     \\ 
Regularized Dilated Optimistic Mirror Descent (Reg-DOMD)                      & Full Information                                                               &      \citet{DBLP:conf/iclr/LiuOYZ23-power-reg}     \\ 
Regularized Counterfactual Regret Minimization (Reg-CFR)                      & Full Information                                                               &      \citet{DBLP:conf/iclr/LiuOYZ23-power-reg}     \\ 
Magnetic Mirror Descent (MMD)                                 & Full Information / \newline Outcome Sampling &     \citet{DBLP:conf/iclr/SokotaDKLLMBK23-MMD}      \\
Q-Function Based Regret Minimization (QFR)                    & Full Information /\newline Outcome Sampling  &       \citet{QFR}    \\ 
Clairvoyant Mirror Descent (CMD)                              & Full Information                                                               &      \citet{wibisono2022alternating-clairvoyant}     \\ 
\bottomrule
\end{tabular}}
\caption{The baseline algorithms implemented by \LiteEFG.}
\label{table:baselines}
\end{table}

\section{Benchmark}

\begin{figure}[thp]
    \centering
    \begin{tikzpicture}
        \node[text width=8.14cm,] at (-4.1,0) {~\\[2mm]\includegraphics[scale=.56]{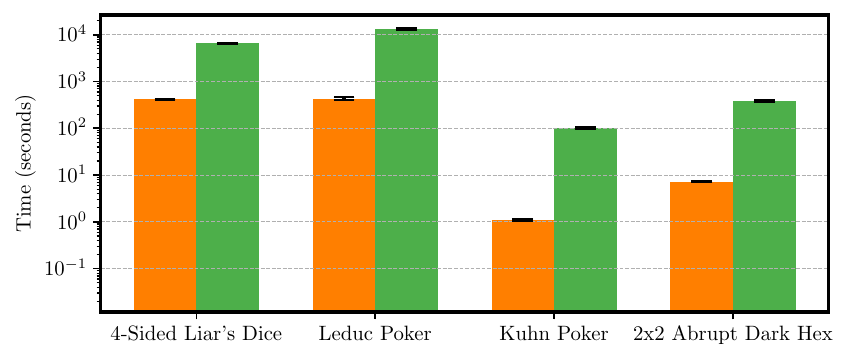}};
        \node[text width=8.14cm,] at (+4.1,0) {~\\[2mm]\includegraphics[scale=.56]{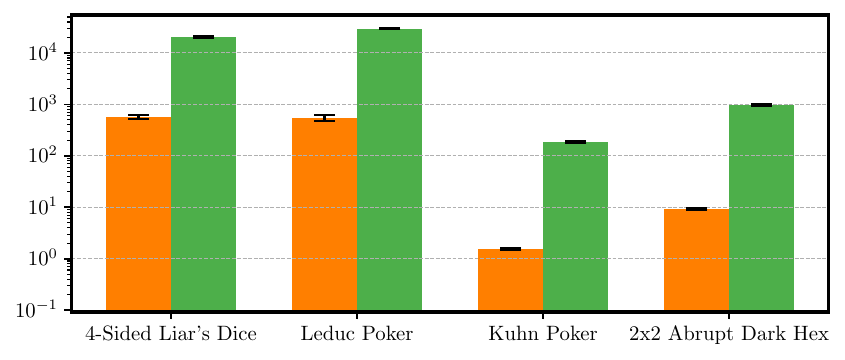}};
        
        \node[fill=white,text=black,text width=8.14cm,align=center] at (-4.1, 1.78) {~};
        \node[fill=white,text=black,text width=8.14cm,align=center] at (4.1, 1.78) {~};
        \node[fill=white,text=black,rounded corners=2mm,text width=8.14cm,align=center] at (-4.1, 1.8) {\small CFR \citep{DBLP:conf/nips/ZinkevichJBP07-CFR}};
        \node[fill=white,text=black,rounded corners=2mm,text width=8.14cm,align=center] at (4.1, 1.8) {\small CFR+ \citep{DBLP:conf/ijcai/TammelinBJB15-CFR+}};
        \node at (0,-2.2) {\includegraphics[scale=0.7]{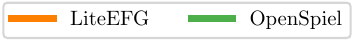}};
    \end{tikzpicture}
    \vspace{-6mm}
    \caption{The running time of CFR and CFR+ in 4 benchmark games, Liar's Dice, Leduc Poker, Kuhn Poker, and Dark Hex. Each algorithm is executed for 100 times in each game, and for each execution, the algorithm lasts for 100,000 iterations. From the figure, \LiteEFG is about 100$\times$ faster than OpenSpiel.}
    \label{fig:experiments}
\end{figure}
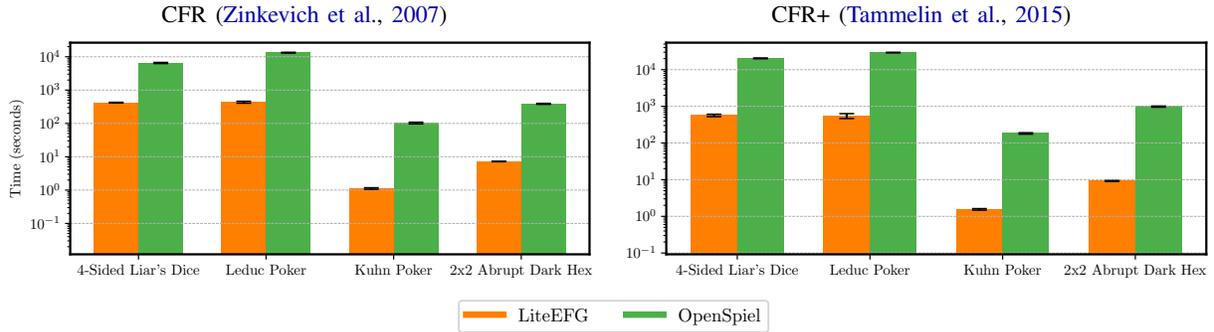

In this section, we will compare the performance of CFR over \LiteEFG and OpenSpiel. All experiments are computed on {\tt Intel(R) Xeon(R) Platinum 8260 CPU @ 2.40GHz}. We compare the performance of the baseline algorithm CFR \citep{DBLP:conf/nips/ZinkevichJBP07-CFR} and CFR+ \citep{DBLP:conf/ijcai/TammelinBJB15-CFR+} in four classical benchmark games, Liar's Dice, Leduc Poker \citep{Leduc}, Kuhn Poker \citep{Kuhn}, and Dark Hex. The results are shown in \Cref{fig:experiments}. To make a fair comparison, we directly call the official C++ implementation of CFR / CFR+ in OpenSpiel by {\tt open\_spiel.python.algorithms.cfr.\_CFRSolver}.

We can see that in \Cref{fig:experiments}, \LiteEFG provides over $100\times$ acceleration in both games, compared to OpenSpiel. In OpenSpiel, running CFR+ for 100,000 iterations takes about 8 hours in Leduc Poker, while it takes less than 10 minutes with \LiteEFG. In the regime of learning in EFGs, the algorithms usually require extensive hyper-parameter search to enhance the performance \citep{DBLP:conf/nips/LeeKL21-EFG-last-iterate, DBLP:conf/iclr/LiuOYZ23-power-reg, DBLP:conf/iclr/SokotaDKLLMBK23-MMD}, which further enlarges the gap between \LiteEFG and OpenSpiel. 

\section{Conclusion}

In this paper, we propose the new computation framework \LiteEFG for solving EFGs. It is more computationally efficient compared to previous work. Moreover, users can avoid handling the complex imperfect-information structure of EFGs by using \LiteEFG, since the library will automatically process the data flow between decision nodes with imperfect information and real nodes in the game tree. Therefore, \LiteEFG would benefit researchers by improving the efficiency of both implementation of the code and computation for algorithms in EFGs.

\bibliographystyle{plainnat}
\bibliography{main}

\end{document}